\documentclass[aps, showpacs, prl, twocolumn, 10pt]{revtex4}

\usepackage{amssymb, amsmath} 

\usepackage{bbm} 

\usepackage[english]{babel}
\usepackage{slashed}

\usepackage{graphicx}
\usepackage{color}

\usepackage[bookmarks,breaklinks,plainpages=false,pdfpagelabels]{hyperref}  
	\hypersetup{linkcolor=red,citecolor=blue,filecolor=dullmagenta,urlcolor=darkblue} 

\smallskipamount
\definecolor{grey}{rgb}{0.4,0.4,0.4}
\definecolor{dullmagenta}{rgb}{0.4,0,0.4}
\definecolor{darkblue}{rgb}{0,0,0.4}
\definecolor{orange}{rgb}{1,0.5,0}
\definecolor{lightbrown}{rgb}{0.75,0.5,0.25}
\definecolor{tan}{cmyk}{0.14,0.42,0.56,0}
\definecolor{djunglegreen}{cmyk}{0.99,0,0.52,0}
\definecolor{lightgreen}{rgb}{0,1,0}
\definecolor{olivegreen}{cmyk}{0.64,0,0.95,0.40}
\definecolor{midgreen}{rgb}{0.0,0.675,0.0}

\newcommand{\vs}{\vspace}

\renewcommand{\.}{\hspace{0.5mm}}

\newcommand{\Ecal}{\ensuremath{\mathcal{E}}}

\newcommand{\Mcal}{\ensuremath{\mathcal{M}}}

\newcommand{\Ocal}{\ensuremath{\mathcal{O}}}

\newcommand{\Scal}{\ensuremath{\mathcal{S}}}

\renewcommand{\d}{\ensuremath{\mathrm{d}}}

\newcommand{\defas}{\mathrel{\mathop :}=} 

\newcommand{\cf}{c.f.}

\makeatletter
  \def\ifundefined#1{\expandafter\ifx\csname#1\endcsname\relax}
  \ifundefined{default@color}
    \let\default@color=\current@color
  \fi
\makeatother

\newcommand{\beq}{\begin{equation}}
\newcommand{\eeq}{\end{equation}}
\newcommand{\bea}{\begin{eqnarray}}
\newcommand{\beas}{\begin{eqnarray*}}

\newcommand{\eea}{\end{eqnarray}}
\newcommand{\eeas}{\end{eqnarray*}}

\newcommand{\m}{\mathrm{m}}

\begin{document}

\title{Island of Stability for Consistent Deformations of Einstein's Gravity}

\author{Felix Berkhahn$^a$}\email{felix.berkhahn@physik.lmu.de}
\author{Dennis D.~Dietrich$^{bc}$}\email{dietrich@cp3.sdu.dk}
\author{Stefan Hofmann$^a$}\email{stefan.hofmann@physik.lmu.de}
\author{Florian K{\"u}hnel$^a$}\email{florian.kuehnel@physik.lmu.de}
\author{Parvin Moyassari$^a$}\email{parvin.moyassari@physik.lmu.de}
\affiliation{
$^a$Arnold Sommerfeld Center, Ludwig-Maximilians University, Theresienstr.~37, 80333 M{\"u}nchen, Germany\\
$^b$CP$^\mathit{3}$-Origins, Centre for Particle Physics Phenomenology, University of Southern Denmark, Campusvej 55, 5230 Odense M, Denmark\\
$^c$\mbox{Institute for Theoretical Physics, Goethe University, Max-von-Laue-Str.~1, 60438 Frankfurt am Main, Germany}
}

\date{\today}

\begin{abstract}
We construct deformations of general relativity that are consistent and 
phenomenologically viable, since they respect, in particular,
cosmological backgrounds. These deformations have unique symmetries
in accordance with their Minkowski cousins (Fierz-Pauli theory for massive gravitons)
and incorporate a background curvature induced self-stabilizing mechanism. 
Self-stabilization is essential in order to guarantee hyperbolic evolution in
and unitarity of the covariantized theory, as well as the deformation's uniqueness. 
We show that the deformation's parameter space contains islands
of absolute stability that are persistent through the entire cosmic evolution. 
\end{abstract}


\maketitle


{\it Introduction \& Overview}{\;---}
Cosmology encompasses the relativistic domain of gravity 
and allows to investigate the rigidity of Einstein's theory.
Consistent deformations of general relativity have been investigated at the
level of linear perturbations on a frozen Minkowski background. 
Fierz and Pauli showed that this system allows for a unique 
deformation satisfying all stability requirements for the prize of
introducing new degrees of freedom corresponding to additional
helicities of a massive graviton.

The new degrees of freedom consistently violate the principle of 
equivalence by constituting a source filter that decreases the vacuum's 
weight on space-time in a technical natural way, albeit due to a delicate 
mass term. This offers a dynamical mechanism 
to address the staggering conflict between naive but educated expectations
for our vacuum's energy density and a plethora of data probing 
the background expansion history and the evolution of density perturbations
in our Universe at various epochs.

In this Letter we covariantize the Fierz-Pauli mass term to a deformation 
that is capable of coexisting with generic cosmological backgrounds.
Requiring hyperbolic evolution and unitarity allows the covariantized theory
to inherit the uniqueness property of its Minkowski cousin (Fierz-Pauli theory).
We show that absolute stability is guaranteed via a background induced 
self-sourcing (feedback) mechanism that is already operational at the linear level. 

For realistic cosmological backgrounds the deformation of Einstein's theory
is characterized by three parameters and its symmetries agree with
those of the Fierz-Pauli mass term. The parameter space features a 
multi-facetted stability dynamics: It includes strictly forbidden regions,
regions that are consistent but challenged through strong coupling, and
parameter islands that support absolutely stable deformations.

{\it Framework}{\;---}
At the linear level the leading relevant deformation of general relativity can be written as a field theory for the combination 
\begin{align}
	H_{\mu\nu} 
		&=						h_{\mu\nu} + \nabla_{\!(\mu} A_{\nu)}
								+ \nabla_{\!\mu}\nabla_{\!\nu} \Phi \;.
								\label{eq:H}
\end{align}
Here, $h$, $A$, $\Phi$ are rank-2,1,0 tensors, respectively, under full background diffeomorphisms; round brackets around indices stand for symmetrization. This parametrization corresponds to two successive St{\"u}ckelberg completions and introduces a $U(1)^4\times U(1)$ gauge symmetry among the fields $h$, $A$, $\Phi$. 

The action for $H$ on a spacetime $( M, g_{0} )$ reads 
\begin{equation}
	\Scal [H]
		=						\frac{ 1 }{ 2 } \int_M\!\d^{4} x\; \sqrt{ | g_{0} | }\,
								H^{\!\sf T}_{}
								\left[
								{\Ecal}( g_{0}, \nabla )
								+ \Mcal( g_{0} )
								\right] H
								\, .
								\label{eq:S-schematic}
\end{equation}
Here, $g_{0}$ denotes the background metric, ${\Ecal}( g_{0}, \nabla )$ is the kinetic operator for $h$, obtained from linearizing the Einstein tensor around the background $g_{0}$, and $\nabla$ denotes the $g_{0}$-compatible covariant derivative. Note that the St{\"u}ckelberg combination in \eqref{eq:H} is effectively an element of $\ker\!\big[ {\Ecal}( g_{0}, \nabla ) \big]$, once sources have been supplied. The deformation operator is denoted by $\Mcal( g_{0} )$ and is, at this stage, of second adiabatic order (given by the number
of derivatives acting on the background metric) barring parameters with inverse mass dimension.

{\it Uniqueness}{\;---}
The Goldstone-St{\"u}ckelberg field $\Phi$ enters the gauge invariant combination $H$ with two derivatives and, therefore, the action \eqref{eq:S-schematic} with four derivatives. Without further restricting $\Mcal( g_{0} )$, the short distance behavior of the deformation would be governed by a higher-derivative theory that violates unitarity. A similar conclusion holds for the field $A$. Now, the necessary and sufficient condition on $\Mcal( g_{0} )$ to yield only second order equations of motion for $\Phi$ and $A$ is 
%
$	\Mcal^{\mu\nu\alpha\beta}
		=						- \Mcal^{\alpha\nu\mu\beta}
$,								
%
in addition to the previously-discussed symmetries.

As a result, to second adiabatic order, the deformation operator can be expanded uniquely as 
\begin{align} \label{eq:M_verygeneral}
	\Mcal^{\mu\nu\alpha\beta}
		&=						\left(
								m_0^{2}
								+ \alpha R_{0}
								\right)\!
								g_{0}^{\; \mu[ \nu}\.g_{0}^{\; \beta]\alpha}
								\\[1.5mm]
		&+						\beta\Big(\.
								R_{0}^{\;\mu [\nu}\.g_{0}^{\;\beta]\alpha}
								+\.R_{0}^{\; \alpha[ \beta}\.g_{0}^{\;\nu] \mu}
								\Big)
								+ \gamma\.R_{0}^{\;\mu\alpha\nu\beta}\notag
								\;,
\end{align}
where the subscript $0$ indicates a background quantity, $\alpha, \beta, \gamma$ are real dimensionless parameters, and square brackets around indices stand for anti-symmetrization. Including terms of higher adiabatic order requires introducing further parameters with appropriate inverse mass dimension to compensate for the additional derivatives acting on $g_{0}$.

To lowest adiabatic order ($\alpha, \beta, \gamma=0$), $\Mcal$ coincides 
with a na{\"i}vely covariantized Fierz-Pauli term and reduces precisely to the well-known Fierz-Pauli mass deformation on the Minkowski background \cite{Fierz:1939ix}.


{\it Stability Analysis}{\;---}
The stability analysis only requires to determine the roots of the determinant of the kinetic operator, \cf~\eqref{eq:S-schematic}, which signal the saturation of the stability or unitarity bounds \cite{Berkhahn:2010hc}, respectively.
In order to calculate the determinant of the kinetic operator it is useful to completely fix the gauge to $h_{0 \mu} = 0$ and $A_{0} = 0$. The saturation of the unitary bound is marked by the zero crossing of the coefficient in front of the highest power in the temporal component of the momentum, which here is ${k_{0}}^{20}_{}$. 
The stability bound is determined by the zero crossing of the coefficient in front of the highest power in the spatial components of the momentum, which here is $( k_{0}\.k_{j} )^{10}_{}$. 

In general, we distinguish the following four cases. 
{\it Case 1}: Both bounds are satisfied on the entire spacetime. Hence, the deformation is well-defined at the perturbative level. 
{\it Case 2}: 
Regions that support both bounds are separated from areas where the unitarity bound is violated by regions in which the stability bound is violated. 
This situation is called `self-protected' \cite{Berkhahn:2010hc}, and which is in accordance with the classicalization mechanism \cite{Dvali:2010bf}.
{\it Case 3}: There are spacetime regions on which both bounds are satisfied, and these regions have a common border with regions where unitarity is violated. In this case, the theory must be dismissed, as the unitarity violation diagnosed at the linear level cannot be cured by a nonlinear completion. {\it Case 4}: The theory is unstable or unitarity violating in the observer's spacetime region.

Let us specialize to a Friedmann spacetime, $g_{0} = {\rm diag}( -1, a^{2}, a^{2}, a^{2} )$, 
where $a = a( t )$ is the scale factor. 
{\it Case 2} corresponds to a situation where a healthy region at late times $t$ ("today") is preceded by a stability violating one, which always separates the former from a potentially present but even earlier unitarity violating region. Correspondingly, in {\it Case 3} the healthy is preceded by a unitarity violating 
regime without an intermediate unstable phase. For $m_{0} \neq 0$ but $\alpha, \beta, \gamma = 0$ 
this case never occurs \cite{Berkhahn:2010hc}. 
\begin{figure}[t]
	\begin{center} 
	\scalebox{1.8}{
			\includegraphics[width=4cm]{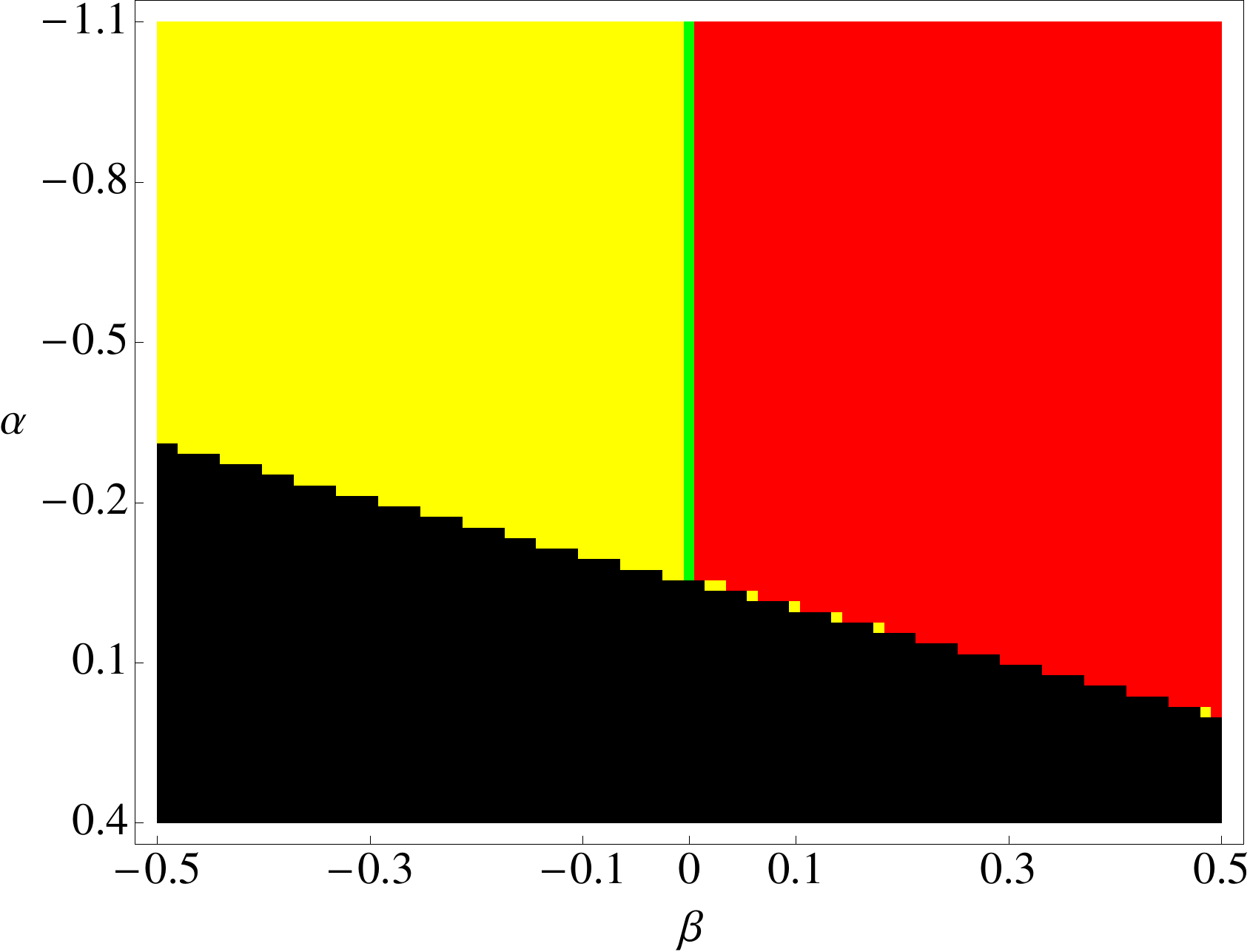}}
		\caption{	
				Parameter plot in the $\alpha$-$\beta$ plane, for $m_{0} = 0$. 
				The green (top center) line corresponds to {\it Case 1}, 
				the yellow (top left) region to {\it Case 2}, 
				the red (right) to {\it Case 3}, and the black (bottom) region to {\it Case 4}. 
				(For $m_{0} = H_{0}$ the plot looks essentially the same.)
				\vs{-5mm}
				}
		\label{fig:paraspace-m=H_0}
	\end{center}
\end{figure}
For the Friedmann metric, the Riemann tensor can be expressed through the Ricci tensor, the Ricci scalar, and the background metric. Thus, without loss of generality, we can set $\gamma$ to zero in \eqref{eq:M_verygeneral}.
Remarkably, then, and {\sl this is our main result}, an appropriate choice of $\alpha$ and $\beta$ makes the deformation absolutely stable, corresponding to {\it Case 1}. (See Fig.~\ref{fig:paraspace-m=H_0}.) 

To proceed, we parametrize time with the scale factor $a$, which is determined using
the observed mixture of matter and radiation densities, and the cosmological constant \cite{Larson:2010gs}
as sources for the cosmological concordance model. 
Parametrizing time with the scale factor gives the unitarity and stability bound as a polynomial in 
$a$. 

We have shown analytically that the interplay between both bounds results in $\beta = 0$ as a necessary condition for obtaining absolute stability over the entire cosmological expansion history.
It turns out that the radiation dominated epoch restricts the stability dynamics considerably. 
Moreover, we find as a condition for absolute stability $\alpha < \alpha_{\rm max} < 0$, where $\alpha_{\rm max}$ depends on the precise mixture of cosmological sources. The expression for 
$\alpha_{\rm max}$ is rather involved and will be presented elsewhere. 
In addition, there is an isolated point of absolute stability in parameter space,
given by $\alpha = m_{0}^{2} / (48\, \Omega_{\Lambda})$, $\beta = 0$, with $\Omega_{\Lambda}$ denoting the current relative density parameter of the cosmological constant.


{\it Covariantized deformation parameter}{\;---}
Our results show that the models with
\begin{align}
	\Mcal^{\mu\nu\alpha\beta}( g_{0} )
		&=						\big(
								m_0^{2}
								+ \alpha\.R_{0}
								\big) \left(
								g_{0}^{\; \mu \nu}\.g_{0}^{\; \alpha\beta} - 
								g_{0}^{\; \mu \beta}\.g_{0}^{\; \alpha\nu}
								\right)
								\label{eq:M_min}
\end{align}	
yield a completely stable theory if $\alpha<\alpha_{\rm max}<0$. 
It is tempting, albeit not quite correct, to think of \eqref{eq:M_min} as 
a 'running mass' deformation. 

The leading short-distance behavior of the minimal deformation \eqref{eq:M_min} is captured by the action
\begin{align}
	\Scal[ \Phi ] 
		&\simeq					\int \d^4 x \; \sqrt{|g_{0}|} \;
								\Phi \; \Ocal^{\mu\nu} \nabla_{\!\mu}\nabla_{\!\nu} \; \Phi
								\; ,
								\label{eq:sd}
\end{align}
with the pseudo metric 
\begin{align} 
	&\Ocal^{\mu\nu}
		\defas					\Bigg[\.
								4\.\frac{ \big( \nabla^{\alpha} \m^{2} \big)
								\big( \nabla_{\mspace{-5mu}\alpha} \m^{2} \big) }{ \m^{2} }
								- 3\.\m^{4}
								- 2\.\big( \Box\mspace{2mu}\m^{2} \big)
								\Bigg]\.
								g_0^{\; \mu\nu}
								\notag
								\\[2mm]
		&						+ 2
								\Bigg[\.
								\m^{2}\.R_{0}^{\;\mu \nu}
								+ \big( \nabla^{\mu} \nabla^{\nu} \m^{2} \big)\!
								- 2\.\frac{ \big( \nabla^{\mu} \m^{2} \big)
								\big( \nabla^{\nu} \m^{2} \big) }{ \m^{2} }
								\Bigg]
								,
								\label{eq:O_phi}
\end{align}
where $\m^{2}=m_0^2+\alpha R_0$.
For a Friedmann spacetime, the action \eqref{eq:sd} can be brought into the form
\begin{align}
	\Scal[\Phi]
		&\simeq					\int \d^4 x \;
								\left[
								A( t )\big(\dot{\Phi}\big)^{2}
								+ B( t ) \left({\bf\nabla}\Phi\right)^{2}
								\right]
								,
								\label{eq:frw_action}
\end{align}
with known functions $A( t )$ and $B( t )$. A violation of unitarity/stability is heralded by a sign
flip of $A$/$B$. 

\begin{figure}[t]
	\begin{center} 
	 \scalebox{0.24}{
		\includegraphics{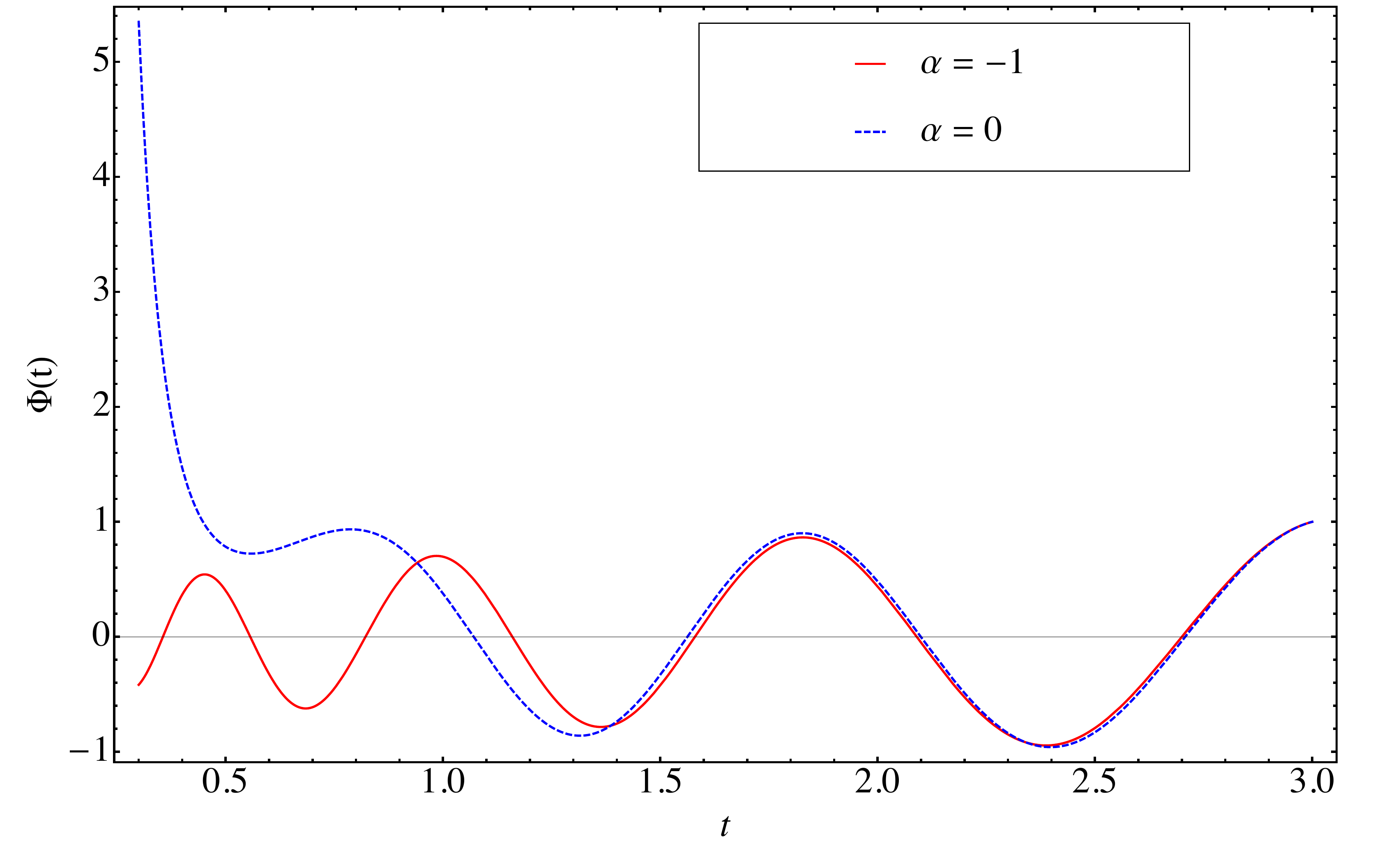}
		}
				\caption{
				$\Phi( t )$ on a purely matter dominated background,
				$m_{0} = H_{0}$, and $\beta, \gamma = 0$. 
				The dashed blue curve corresponds to $\alpha = 0$, 
				while the solid red line corresponds
				to the absolutely stable situation with $\alpha = -1$.
				\vs{-5mm}
				}	
		\label{fig:phi}
	\end{center}
\end{figure}

Fig.~\ref{fig:phi} shows the dominant short distance degree of freedom $\Phi( t )$ for the specified deformation parameters, corresponding to a na{\"i}vely covariantized Fierz-Pauli term and the covariantized deformation parameter model, respectively, as an example for {\it Case 1}. As can be seen, a static deformation parameter ($\alpha = 0$) yields a theory that is self-protected against unitarity violations by a strong coupling regime at the linear level, but is unfit from a phenomenological point of view. The running deformation parameter results in a model that is absolutely stable and, hence, potentially phenomenologically viable.\\

{\it Phenomenology and Applications}{\;---}
The phenomenology of our theory in the solar system will be the same as the one of standard massive gravity with mass $m_0$ \cite{Dvali:2002vf}, since in this environment we have $R=0$. For example, mercurys perihelion advance per orbit $\delta \phi$ due to the gravity modification will be given by $\delta \phi = \pi r \frac{d}{dr}\left( r^2 \frac{d}{dr}(r^{-1} \epsilon) \right)$, with $\epsilon = e^{-m_0 r}$ ($r$ being the mean distance of mercury to the sun). The theory with $m_0 = 0$ will yield the same phenomenological predictions on the linear level as general relativity, and is thus unconstrained from solar system experiments. 

However, we will get a modification on cosmological scales where the Friedmann expansion applies. Taking as a reasonable scale $-\alpha \sim 1$, the effective graviton mass will automatically be in the interesting cosmological domain $m^2 \sim H^2$. In our theory, this scale is naturally set by a dynamical mechanism and does not have to be put in by hand. 

As an application, we consider gravitational waves on a de Sitter background with cosmological constant $\Lambda$, the equation of motion for the rank-2 
tensor is given by 
\begin{eqnarray} 
	\label{eqn:degr}
	\hat{\mathcal{E}}^{\; \; \; \alpha \beta}_{\mu \nu}(g^0,\nabla) h_{\alpha \beta} - 
	\frac{1}{3} \Lambda \left( h_{\mu \nu} + \frac{1}{2} g_{ \mu \nu}^0 h \right) + \nonumber \\ 
	- \left( m_0^2 + 4 \alpha \Lambda \right) \left( h_{\mu \nu} - 
	g^0_{\mu \nu} h \right) = \delta T_{\mu \nu} \; ,
\end{eqnarray}
where $\hat{\mathcal{E}}$ is the part of the linearized Einstein tensor containing 
covariant derivatives acting on $h$, and $\delta T$ denotes the perturbation of 
a covariantly conserved background source. In our case, $\delta T=\delta\Lambda \, g^0$,
where $\delta \Lambda$ is an additional de Sitter source. 
Clearly, $h=C\, g^0$ is a solution of (\ref{eqn:degr}), provided
$C=-\delta\Lambda / [(1-12\alpha)\Lambda -3 m_0^{\;2}]$, resulting 
in the total metric field $g=(1+C)\.g^0$. The total curvature is related to the 
background curvature as $R=R_0/(1+C)\approx (1-C) R_0$. In general relativity, 
this is $R=4\.(\Lambda + \delta \Lambda)$, 
as expected. In contrast, if the deformation is operative,
then $|C|$ can be smaller (note that $\alpha < 0$ on the stable island),
and the resulting curvature can be smaller as compared to the previous case.
The effect of $\delta \Lambda$ on the curvature is partially degravitated \cite{Dvali:2007kt}. 

As a second application, let us calculate the gravitational potential of a point 
particle with mass $\mu$ on a de Sitter background. 
Parametrizing the scale factor as $a( t )\equiv {\rm exp}(\sqrt{\Lambda/3}\; t)$
and defining $\hat{h}_{00}=a^{7/2}( t ) \; h_{00}$, we have
\begin{equation}
\label{eq}
\left(\partial_t^2 +\omega(k_{\rm p})\right) \hat{h}_{00}=4 \sqrt{a}\mu/3 \;,
\end{equation}
where
\begin{equation}
\omega^2(k_{\rm p})=k^2_{\rm p}( t )+\m^2-3\Lambda/4,
\end{equation}
and $k_{\rm p}$ denotes the physical wavenumber, $k_{\rm p}( t )\equiv k/a( t )$
in terms of the comoving wavenumber $k$. 

Using the WKB approximation, we recover the Yukawa potential at early times, $t\ll1/\sqrt{\Lambda/3}$,
\begin{equation}
V(r_{\rm p})\propto \frac{\mu \, \exp{(-\m\,r_{\rm p}( t ))}}{4\.\pi\.r_{\rm p}( t )} \;, 
\end{equation}
with $r_{\rm p}$ denoting the physical distance, $r_{\rm p}( t )\equiv a( t )\. r$
in terms of the comoving distance $r$.

At late times, $t\sim 1/\sqrt{\Lambda/3}$, $\omega$ becomes $k_{\rm p}$ independent 
and the gravitational potential becomes $V(r_{\rm p})\sim \delta(r_{\rm p})$.
This shows that in the deformed theory the effective interaction range of two point particles 
generically is much smaller than their physical distance at late times, see Fig.~3. 

\begin{figure}[t]
 \begin{center}
 
 \scalebox{0.24}{ \includegraphics{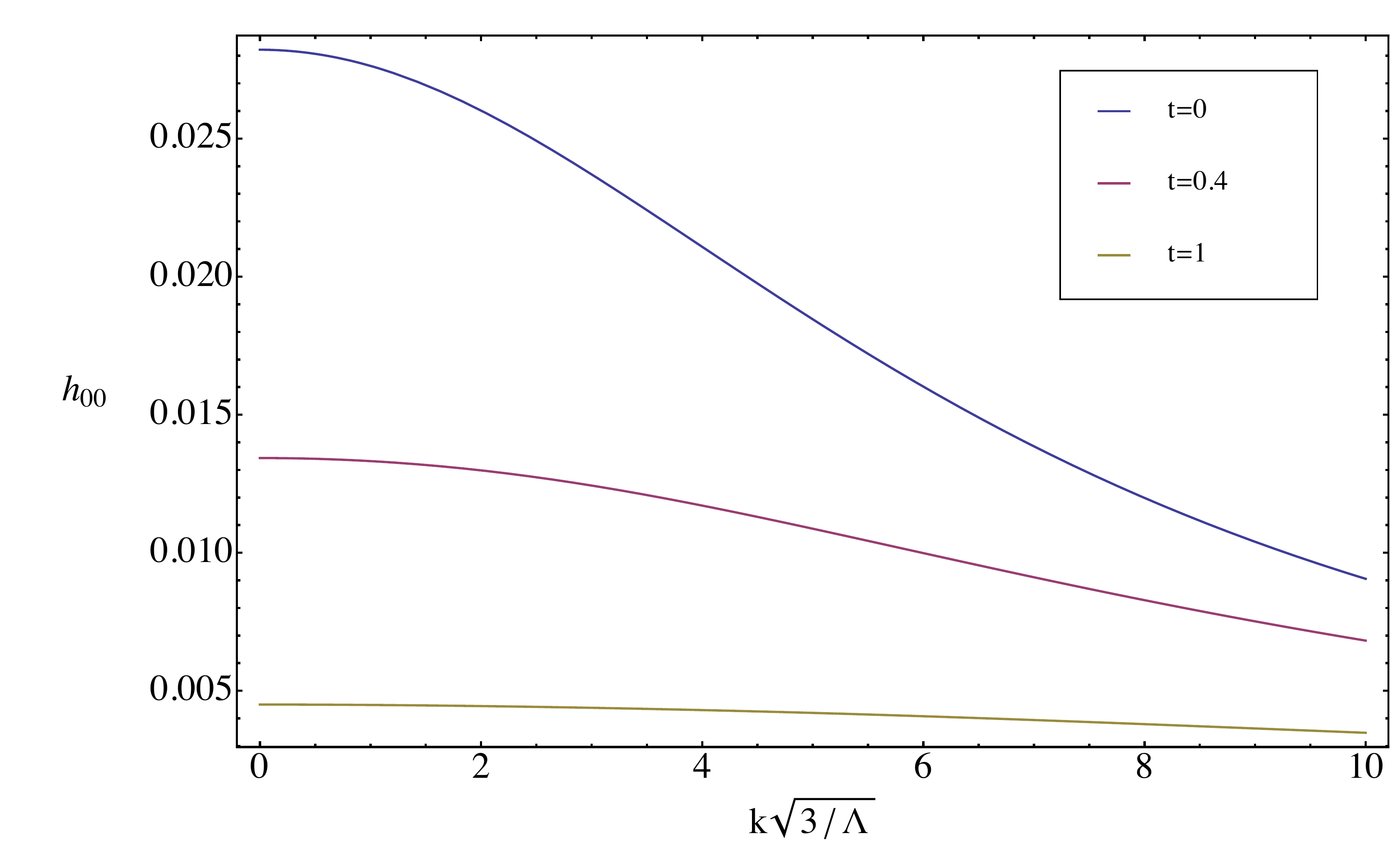}}
 \end{center}
 \caption{The evolution of $h_{00}(k)$ at different times for $\mu=\sqrt{\Lambda/3}=1$ and $\m=5$. 
The blue line (upper one) is at $t=0$, the red line (middle one) at $t=0.4$ and yellow line (lower one) at 
$t=1$.}
 \end{figure}

{\it Conclusion \& Summary}{\;---}
In this letter we have constructed a unique deformation of gravity
corresponding to a covariantized Fierz-Pauli theory for massive gravitons 
that posses islands of absolute stability in its parameter space
over the entire cosmic expansion history. The uniqueness property is 
a legacy of the deformation's Minkowski cousin when requiring classical
stability and unitarity. We are hopeful that this deformation represents
an exciting window of opportunity for studying consistent modifications
of gravity on the largest observable distances. 

Certainly, the very important question about a possible nonlinear completion
of this unique deformation remains. However, we achieved a consistent 
covariantization of Fierz-Pauli theory on realistic cosmological backgrounds that 
is kept healthy via a background induced self-stabilization mechanism. 
We are currently working on a nonlinear completion. A nonlinear completion
of the hard mass deformation ($\alpha = 0$) has been recently constructed in a 
different framework \cite{deRham:2010kj}.


{\it Acknowledgments}{\;---}
It is a pleasure to thank Eugeny Babichev, Andrei Barvinsky, Yannis Burnier, Gia Dvali, Fawad Hassan, Justin Khoury, Michael Kopp, Slava Mukhanov, Kerstin Paech, as well as Florian Niedermann for helpful discussions, and {\sc nordita} for the stimulating workshop {\it The Return of de Sitter}. The work of SH and FK was
supported by the DFG cluster of excellence 'Origin and Structure of the Universe'.
The work of FB and SH was supported by TRR 33 'The Dark Universe'.
The work of PM was supported by the Alexander von Humboldt Foundation.
\vs{-2mm}


\end{document}